\documentclass[prl,aps,showpacs,twocolumn]{revtex4}
\usepackage{graphicx}
\usepackage{amsmath,amssymb,revsymb}
\usepackage[colorlinks=true,citecolor=blue,linkcolor=blue]{hyperref}
\begin{document}

\title{Universal three-body physics for fermionic dipoles}

\author{Yujun Wang}
\author{J. P. D'Incao}
\author{Chris H. Greene}
\affiliation{Department of Physics and JILA, University of Colorado, Boulder, Colorado, 80309-0440, USA}

\begin{abstract}
A study of the universal physics for three oriented fermionic dipoles in the hyperspherical adiabatic representation 
predicts a single long-lived three-dipole state, which exists in only one three-body symmetry, 
should form near a two-dipole resonance. Our analysis reveals the spatial configuration of the universal state, and the scaling 
of its binding energy and lifetime with the strength of the dipolar interaction.
In addition, three-body recombination of fermionic dipoles is found to be important even at ultracold energies. 
An additional finding is that an effective long-range repulsion arises between a dipole and a dipolar dimer 
that is tunable via dipolar interactions.
\end{abstract}
\maketitle

The recent advances in producing ground-state ultracold dipolar molecules~\cite{DipoleGnd} have sparked great interest in the novel
phases that could be experimentally accessible~\cite{Baranov}. 
With the ability to control the dipolar interaction by an external electric field~\cite{DipoleReact}, 
studies of dipolar molecules could stimulate the discovery of intriguing phenomena across a broad range
of disciplines, from condensed matter physics to ultracold chemistry. 
In particular, the effects of the long-range anisotropic dipolar interaction on ultracold physics 
has already been studied for three bosonic dipoles~\cite{Ticknor,Wang}.
A recent development has shown~\cite{Wang} that the Efimov effect~\cite{Efimov} is more favorable for three bosonic dipoles although 
the dipolar interaction has ingredients that could conceivably have destroyed it.

Stimulated by their ability to mimic a variety of exotic physical systems from superfluids to neutron stars,
quantum degenerate fermi gases with nondipolar interactions have been studied extensively in recent years~\cite{FermiGas}.
In contrast to bosonic gases whose lifetime is mainly determined by three-body recombination~\cite{EsryRecomb}, 
fermionic gases exhibit extraordinary stability against three-body inelastic collisions except 
when the two-body interaction is tuned near a $p$-wave resonance~\cite{Suno,Castin}. Moreover, three-body physics for identical 
fermions is expected to be nonuniversal, i.e., short-range-dependent~\cite{Suno,DIncao}. Consequently, three-body physics for 
identical fermions have to some extent been regarded as less exciting and/or difficult to observe. 
Nevertheless, the influence of dipolar interactions on the universal behavior of bosons triggers new interest in the 
study of universality for fermionic dipoles, and it is becoming increasingly important to understand such physics in order to guide ongoing 
experiments on ground-state fermionic dipolar molecules~\cite{DipoleReact}.

In this Letter, the behavior of three identical fermionic dipoles is demonstrated to be universal, thus differing fundamentally 
from the nondipolar cases. 
In particular, near a dipole-dipole resonance (i.e., where the two-dipole binding energy $E_{2d}\rightarrow 0$), 
a single long-lived, universal three-dipole state is formed. Its size and lifetime 
both grow with stronger dipolar interactions. The stability of the fermionic dipolar gases is assessed through  
numerical calculations of the rate coefficient $K_3$ for three-dipole recombination, 
$D+D+D\rightarrow D_2+D$, which is the dominant expected loss mechanism for 
nonreactive ground-state dipolar molecules~\cite{Hutson}. In fact, we have found that $K_3$ grows rapidly 
with the strength of the dipolar interactions, and it is strongly enhanced near a dipole-dipole resonance.
This study also shows evidence for a long-range, dipolar interaction-dependent repulsion that suppresses $D_2+D$ relaxation collisions when
$D_2$ is deeply-bound. 

Consider the three-dipole Schr{\"o}dinger equation in hyperspherical coordinates (in atomic units): 
\begin{equation}
\left(-\frac{1}{2\mu}\frac{\partial^2}{\partial R^2}+H_{ad}
\right)\psi=E\psi.
\label{SchHyper}
\end{equation}
where $\psi=R^{5/2}\Psi$ is the scaled three-body wavefunction, $\mu$=$m/\sqrt{3}$ is the three-body reduced mass 
for identical particles with mass $m$.  
The adiabatic $H_{ad}$ is 
\begin{equation}
H_{ad}=\frac{{\Lambda}^2(\Omega)}{2\mu R^2}+\frac{15}{8\mu R^2}+V,
\end{equation}
where $\Lambda^2$ is the usual grand angular momentum operator~\cite{SunoHe}, and $\Omega$ are the hyperangles representing
positions of the three particles at fixed hyperradius $R$. 
The potential energy $V=v(\vec{r}_{12})+v(\vec{r}_{23})+v(\vec{r}_{23})$ is a pairwise sum of two-dipole interactions:
\begin{equation} 
v(\vec{r})=V_{0}{\rm sech}^2(r/r_{0})+\frac{2d_{\ell}}{m}\frac{1-3(\hat{z}\cdot\hat{r}) ^2}{r^3} f(r).
\label{DipolarInt}
\end{equation}
The ${\rm sech}^2$ term is isotropic and short-range, and it encapsulates the complicated interactions between two polar molecules. 
The second term with a short-range cutoff $f(r)$ is the 
interaction between two dipoles aligned in the $\hat{z}$ direction. The dipole length is defined in terms of the electric 
dipole moment $d_m$ as $d_\ell=m d_m^2/2$.
The universality of our calculations is tested by varying both $d_\ell$ and $V_0$.

Equation~(\ref{SchHyper}) can be routinely solved, either using the adiabatic representation~\cite{SunoHe} or else through 
slow variable discretization (SVD)~\cite{SVD}.
Our study shows it is advantageous to use the adiabatic scheme to treat the asymptotic behavior of three dipoles in scattering calculations, 
while SVD efficiently handles sharp avoided crossings at small distances. 
In both cases, the biggest challenge is to solve the adiabatic eigenvalue equation:
\begin{equation}
H_{ad}\Phi_\nu^{\Pi,M}(R;\Omega)=U_\nu(R)\Phi_\nu^{\Pi,M}(R;\Omega),
\label{Eq:Ad}
\end{equation}
where $U_\nu(R)$ is the adiabatic potential for channel $\nu$, 
$M$ is the space-fixed frame projection of the total angular momentum $J$, 
and $\Phi_\nu^{\Pi,M}(R;\Omega)$ is the corresponding eigenfunction.
For interacting oriented dipoles, the major difficulty in solving Eq.~(\ref{Eq:Ad}) is that 
the total orbital angular momentum $J$ is not conserved. 
The method of Ref.~\cite{Wang} is implemented to solve Eq.~(\ref{Eq:Ad}). 
Briefly, $\Phi_\nu^{\Pi,M}$ is expanded in terms of the Wigner $D$ functions [see Eq.~(4) in Ref.~\cite{Wang}], truncated 
at $J_{\rm max}=13$. Tests of this truncation confirm that it yields adequate convergence. 

\begin{figure}
\includegraphics[scale=0.67]{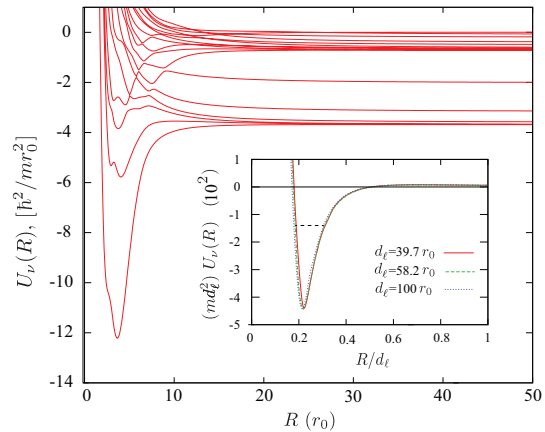}
\caption{
(color online) 
A typical set of adiabatic hyperspherical potentials $U_\nu(R)$ for three fermionic dipoles with $d_\ell/r0=58.2$ and $E_{2d}\rightarrow 0$. 
The inset shows rescaled, diabatized potentials exhibiting universal behavior for a few values of $d_\ell$ at a dipole-dipole resonance.
The horizontal dashed line in the diabatic potential wells indicates the position of the universal three-dipole states.
 }
\label{Fig:BoundPot}
\end{figure}

The nature of three-dipole states near a dipole-dipole resonances emerges from study of the adiabatic potentials 
for $M^\Pi=0^+$. 
Note that the even and the odd $J$ values decouple in this symmetry. 
This symmetry with only odd $J$ is of particular interest because it includes the $J^\Pi=1^+$, the least repulsive partial wave for three 
noninteracting identical fermions~\cite{EsryThresh}. One would also expect the binding of three dipoles is most likely to 
occur at $m_{2d}=0$ dipole-dipole resonance, where $m_{2d}$ is the projection of two-dipole angular momentum along the field direction. 
However, $m_{2d}=0$ is not allowed for $M^\Pi=0^+$ symmetry, 
so the three-dipole system is considered instead near the 
$m_{2d}=1$ resonances. Figure~\ref{Fig:BoundPot} shows the typical behavior of the adiabatic potentials. 
Asymptotically, the three-body continuum potentials behave in the same manner as the nondipolar interaction case, but 
for dipole plus dipolar dimer channels, each threshold has a family of potential curves whose centrifugal barriers carry 
different effective angular momenta due to the $J$-coupling. 

Irregardless of the complicated topology of the adiabatic potentials shown in Fig.~\ref{Fig:BoundPot},  
near a dipole-dipole resonance it is possible to trace a diabatic potential well in the channel closest to the three-body continuum. 
It cuts through multiple sharp avoided crossings with deeply-bound channels, as is visible in the inset of Fig.~\ref{Fig:BoundPot}. 
Remarkably, all these potentials for different $d_\ell$ and for different short-range 
potentials fall on top of each other after a proper scaling. 
A key implication of the scaling of the universal potentials is that the hyperradius of the repulsive 
barrier increases with $d_\ell$, which suppresses decay of the three-dipole state and increases its lifetime 
as the dipolar interaction is increased.  
To quantify the number of three-dipole states and their energies in 
the universal potential wells, we have solved Eq.~(\ref{SchHyper}) by using SVD. 
The energy $E_{3d}$ and the width $\Gamma$ of the three-body states are revealed by 
adding complex absorbing potentials in the lower decaying channels. 
Interestingly, exactly one quasi-bound state is supported by each of the universal potentials.
Table~\ref{Tab:3DipoleEnergies} lists $E_{3d}$ and $\Gamma$ for a few values of
$d_\ell$.
\begin{table}
\begin{ruledtabular}
\begin{tabular}{cccc}
$E_{2d}\rightarrow 0$ &$d_{\ell}$ ($r_0$) & $m d_{\ell}^2E_{3d}$ & $m d_{\ell}^2\Gamma$ \\
\hline
   & 39.7 & 171 & 42 \\
   & 58.2 & 135 & 43 \\
   & 100 & 139 & 17\\ [0.05in] 
\hline
$\Theta\approx 16^\circ$  & $b_l\approx 0.26 d_\ell$ & $b_s\approx 0.14 d_\ell$ & $\Delta\approx 15^\circ$
\end{tabular}
\end{ruledtabular}
\caption{The upper rows: the energies $E_{3d}$ and the widths $\Gamma$ of the universal three-dipole states for different values of $d_{\ell}$. 
The bottom row: the angle $\Theta$ between the three-dipole triangle and field direction, the long and short bond length $b_l$ and $b_s$, and 
the smaller bond angle $\Delta$ from the most probable configuration of the three-dipole states. } 
\label{Tab:3DipoleEnergies}
\end{table}
The energy $E_{3d}$ shown in Table~\ref{Tab:3DipoleEnergies} suggests a universal trend, but the width $\Gamma$ exhibits less universal behavior,
presumably due to the nonuniversal 
couplings to deeply-bound channels. Nevertheless, the $1/d_\ell^2$ suppression in the width implies an increased lifetime 
of the universal three-dipole states as $d_\ell$ increases. 
Our numerical study also shows that the universal three-dipole states slowly become unbound or deeply-bound 
as the system shifts away from a dipole-dipole resonance. 
In a broad region near a dipole-dipole resonance, the properties of the three-dipole states are essentially unchanged. 

\begin{figure}
\includegraphics[scale=0.14]{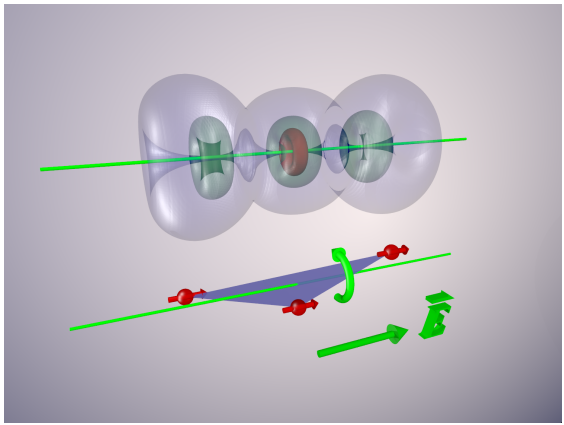}
\caption{
(color online) The upper part: the isosurfaces of the one-dipole probability density $\rho(\vec{r})$ for the universal three-dipole states, 
centered in the space-fixed center of mass frame. By normalizing the peak density to be unity, 
the density of the following surfaces: the red (innermost), three green (middle), and the light blue (outermost) are 0.8, 0.3 and 0.015, respectively.
This crankshaft-resembling structure is a figure of rotation about the field axis.
The lower part: the most probable orientation and geometry of the triangle formed by three dipoles. The green axes show the direction of 
the external electric field.
}
\label{Fig:Isosurface}
\end{figure}

Importantly,  
the interactions of the aligned dipoles break the overall rotational symmetry, reflecting that the three-dipole states have a preferential orientation 
in space. The geometry of these universal states is analyzed by calculating the spatial distribution of the dipoles using the three-body wavefunctions. 
Specifically, the probability density for finding a dipole 
at $\vec{r}$, is
\begin{equation}
\rho(\vec{r})=\frac{1}{3}\langle\Psi\mid \sum_i\delta(\vec{r}-\vec{r}_i) \mid\Psi\rangle,
\end{equation}
where $\delta(\vec{r})$ is the Dirac delta function, $\vec{r}_i$ is the position vector for the $i$'th dipole.
Figure~\ref{Fig:Isosurface} shows the isosurfaces of $\rho(\vec{r})$. The three-dipole states have a strongly preferential  
spatial configuration, as indicated by the triangle in the lower part of the figure. The isosurfaces 
can be viewed as the revolution of the triangle along the field axis. It is also interesting to note that the configuration of the three-dipole states 
are universal, only their overall size scales linearly with $d_\ell$.

The configuration of a three-dipole state shown in Fig.~\ref{Fig:Isosurface} can be understood in terms of the anisotropic nature of the dipolar interactions 
and the symmetry of the state. The three dipoles tend to align in 
a line where the dipolar interactions in Eq.~(\ref{DipolarInt}) are the most attractive, 
but this is forbidden by the odd partial wave symmetry. The state therefore has the configuration of an obtuse, isosceles triangle 
making a small angle with the field direction. The lower part of Table~\ref{Tab:3DipoleEnergies} lists the angle between the plane of the triangle 
and the field direction, the short and long bond lengths, and the smaller bond angle determined from the most probable and also 
the dominant configuration of the three-dipole states. 

The possibility of universal three-dipole states for other symmetries have been explored, in particular the 
three-dipole system for $M^\Pi=0^-$ symmetry with odd $J$'s, which includes the less repulsive partial wave contribution $J^\Pi=1^-$~\cite{EsryThresh}. 
However, no universal three-dipole states have been found near both $m_{2d}=0$ and $m_{2d}=1$ dipole-dipole resonances. The situation is similar for 
$M^\Pi=1^-$ and $1^+$ symmetries, where the adiabatic potentials are found to be more repulsive. The potentials for higher $M$ near 
higher $m_{2d}$ resonances are expected to be more repulsive, suggesting the conclusion that the $M^\Pi=0^+$ three-dipole states presented above are 
the only class of universal states for identical fermions.

The universal three-dipole states affect scattering processes. As $d_\ell$ increases, 
a resonance in three-dipole recombination is expected whenever a three-dipole state passes through the three-body breakup threshold 
before a new two-dipole state becomes bound.
To determine the stability of nonreactive fermionic dipolar gases, however, it is more important to know the general scaling 
of three-dipole recombination with $d_\ell$ and the behavior near a dipole-dipole resonance.
The threshold behavior of the recombination rate $K_3$ for three fermionic dipoles can be obtained from a Wigner law analysis~\cite{EsryThresh}
of the $M^\Pi=0^+$, $J$-odd low-energy-dominant contribution, 
giving
\begin{equation}
K_3=\frac{C_3}{\mu}k^4,
\end{equation}
where $k=\sqrt{2\mu E}$ is the three-body wavenumber. 

For short-range, nondipolar interactions, the coefficient $C_3$ in recombination rate is expected to scale with the scattering volume~\cite{Suno,Castin} 
and the effective range~\cite{Castin}. For dipolar interactions, however, the scattering physics is fundamentally different even in the 
two-body level~\cite{Bohn}. In particular, we have numerically identified a distinct low-energy expansion of the $p$-wave phase shift $\delta$
for two fermionic dipoles. The real part of $\delta$ can be expanded as 
\begin{equation}
{\rm Re}[\delta(k_2)]\approx -a k_2-b k_2^2-V k_2^3\quad (k_2\rightarrow 0),
\label{Eq:PhaseShift}
\end{equation} 
where $k_2$ is the relative two-dipole wavenumber. 
This expansion has a linear term in $k_2$ implying a nonvanishing scattering length $a$~\cite{Bohn} as expected for $r^{-3}$ interaction 
and also a quadratic dependence on $k_2$, both of which are absent for nondipolar interactions. 
But similar to the nondipolar interaction case, the scattering volume $V$ goes through a pole across a dipole-dipole resonance.

\begin{figure}
\includegraphics[scale=0.55]{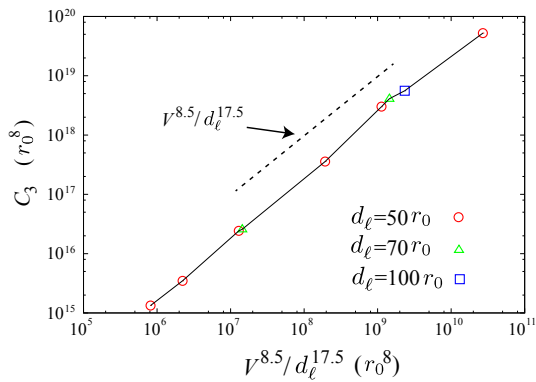}
\caption{
(color online) The dependence of the recombination scaling coefficient $C_3$ on the two-body scattering volume $V$ and 
the dipole length $d_\ell$ near a dipole-dipole resonance. The lines connecting the numerical points just guide the eye.}
\label{Fig:Rates}
\end{figure}

Our scaling laws for three-dipole recombination are based on a calculation of the three-dipole scattering matrix using 
the eigenchannel $R$-matrix propagation method~\cite{Rydberg}.
First, SVD solution of Eq.~(\ref{SchHyper}) obtains the $R$-matrix out to a distance that includes all the sharp-avoided crossings.  
The $R$-matrix is then propagated farther in the adiabatic representation~\cite{SunoHe} in the asymptotic region until $K_3$ 
is fully converged. 
$K_3$ has been calculated for $V>0$ near a dipole-dipole resonance, where the dipolar dimer is weakly bound.
At ultracold energies, our calculation shows that only the lowest three-body continuum channel is important for recombination, and 
$K_3$ is dominated by recombination into the weakly-bound dipole plus dipolar dimer channel having the lowest effective angular momentum barrier asymptotically. 
Figure~\ref{Fig:Rates} shows the scaling coefficient $C_3$ near 
$m_{2d}=1$ dipole-dipole resonances for a few $d_l$ with different short-range interactions. 
The dipole-dipole resonances with higher $m_{2d}$ are not of current interest because of their extremely narrow widths~\cite{Bohn}.
Dimensional analysis has suggested, and our numerical results in Fig.~\ref{Fig:Rates} confirm, that a universal scaling of $C_3$ exists 
\begin{equation}
C_3= \lambda V^{17/2} d_\ell^{-35/2},
\label{Eq:C3}
\end{equation}
where dimensionless $\lambda\approx 2\times 10^{9}$.
This scaling is stronger with $V$ than for nondipolar interactions~\cite{Suno,Castin}.
Although Eq.~(\ref{Eq:C3}) also seems to imply a strong suppression of three-dipole recombination with $d_\ell$, instead a general scaling 
of $C_3\propto d_\ell^8$ is expected away from a dipole-dipole resonance, due to the $d_\ell^3$ scaling of the background value of $V$~\cite{WangUnpub}.
This rapid growth of $C_3$ with $d_\ell$ may be unfortunate for experiments that require stable fermionic dipolar gas in three dimensions, 
since three-dipole recombination can be dramatic even in the ultracold threshold regime ($k d_\ell\ll 1$). 
On the other hand, scenarios exist where three-body recombination has a positive role, 
as a means to produce weakly-bound dipolar dimers that are stable against collision with a free dipole, in close analogy to the technique used in 
ultracold spin-mixed fermions with nondipolar interactions~\cite{MakeDimer}.  

Finally, consider the behavior of deeply-bound dipole plus dipolar dimer channels and their impact on the stability of 
a mixed gas containing both dipoles and dipolar dimers. Similar to the bosonic dipole case~\cite{Wang}, these channels have a centrifugal barrier 
characterized by a non-integer 
effective angular momentum quantum number, and it grows with increasing $d_\ell$. 
Although our numerical study shows that this $d_\ell$-dependence is nonuniversal, 
the effective angular momentum barrier can nevertheless ``protect'' a mixture of dipoles and dipolar dimers 
from collisional decay, and it appears to provide an alternative way to tune interactions in a mixed-species gas.

In summary, a class of universal physics has been identified for three fermionic dipoles. For scattering volumes near a dipole-dipole resonance, one 
universal three-dipole state exists at an energy near the three-dipole breakup threshold. This state has a universal geometry and orientation 
in the external field, with a variable size dependent on $d_l$. Moreover, its long lifetime can be beneficial for  
experimental observation and manipulation. Our study also shows a universal trend of the three-dipole recombination rate, 
which increases rapidly near each dipole-dipole resonance, and grows rapidly with stronger dipolar interactions. 
And the effective repulsion between 
a dipole and a dipolar dimer can be tuned to control collisions in mixed quantum gases.

\begin{acknowledgments}
This work is supported in part by the AFOSR-MURI and by the National Science Foundation. We thank J. L. Bohn for discussions of two-dipole scattering.
\end{acknowledgments}

\end{document}